\documentstyle[prd,aps,epsfig]{revtex}

\newcommand{\bfk}{{\bf k}}

\newcommand{\order}{{\cal O}}
\newcommand{\beq}{\begin{equation}}
\newcommand{\eeq}{\end{equation}}
\newcommand{\beqa}{\begin{eqnarray}}
\newcommand{\eeqa}{\end{eqnarray}}

\newcommand{\lmk}{\left(}
\newcommand{\rmk}{\right)}
\newcommand{\lkk}{\left[}
\newcommand{\rkk}{\right]}

\newcommand{\phidot}{\dot{\phi}}
\newcommand{\rc}{{\cal R}_c}
\newcommand{\Rc}{{\cal R}_c}
\newcommand{\phidots}{\dot{\phi}_s}
\newcommand{\phidotr}{\dot{\phi}_r}
\newcommand{\phidotk}{\dot{\phi}_k}
\newcommand{\phik}{\phi_k}
\newcommand{\vonek}{V'_k}
\newcommand{\vtwok}{V''_k}
\newcommand{\hk}{H_k}

\begin{document}

\title{Curvature perturbation at the local extremum \\ of the inflaton's potential}
\author{Jun'ichi Yokoyama and Shogo Inoue}
\address{ Department
of Earth and Space Science, Graduate School of Science,\\ Osaka
University, Toyonaka 560-0043, Japan }
\date{\today}

\maketitle

\begin{abstract}
The spectrum of curvature perturbation generated during inflation
is studied in the case the inflation-driving scalar
field (inflaton) $\phi$ crosses over its potential extremum.  It is shown
that the nondecaying mode of perturbation has a finite value
 and a proper formula is given.  The result is also extended to more general cases
where  $\ddot{\phi}$ is non-negligible.
\end{abstract}

\maketitle

\vspace*{1cm}

Inflation in the early universe \cite{inf} 
provides a mechanism to generate primordial
density fluctuations out of quantum fluctuations of the
inflation-driving scalar field, {\it inflaton} $\phi$
\cite{HS}, in addition to explain the large-scale homogeneity and
isotropy.  
In the standard inflation models such as new \cite{ni} and
chaotic \cite{ci} inflation, inflation is driven by the potential
energy of the inflaton, $V[\phi]$, as it slowly rolls the potential hill and 
they predict formation of  adiabatic fluctuations with a nearly
scale-invariant spectrum.
More specifically, the amplitude of the curvature perturbation
on comoving hypersurfaces, $\rc$, on
comoving scale $r=2\pi/k$ is given by the formula \cite{sasaki,Mu},
\beqa
   \rc(r) &\cong& \left.\frac{H^2}{2\pi |\dot{\phi}|}\right|_{t_k}
\label{phidot} \\
  &\cong& 
   \left.\frac{3H^3}{2\pi |V'[\phi]|}\right|_{t_k}, \label{vprime}
\eeqa
where a dot denotes time derivative
 and $H$ is the Hubble parameter.  
Here the right-hand-sides should be evaluated at $t=t_k$ when the
relevant scale $k$ left the Hubble radius during inflation.  
In the second approximate equality (\ref{vprime}) we have used the equation of motion 
with the slow-roll approximation, 
\beq
3H\phidot+V'[\phi]=0. \label{slowroll}
\eeq
The reason why (\ref{phidot}) gives an almost
scale-invariant spectrum is that both $H$ and $\dot{\phi}$ change very
slowly during slow-roll inflation.

Recently, however, new classes of inflation models have been proposed
in which $\phi$ is not necessarily slowly rolling during the entire
period of inflation \cite{JY,oi,oiyuragi}. 
Whenever $\ddot{\phi}$ is non-negligible  
in the full equation of motion of the homogeneous part of $\phi$,
\beq
 \ddot{\phi}+3H\phidot+V'[\phi]=0, \label{eqm}
\eeq
the slow-roll approximation (\ref{slowroll}) fails.  There are two
typical cases where this happens. 
One is the case $\phidot$ vanishes.  The proper analysis of the
amplitude of density fluctuation in this case was recently done in \cite{oseto},
where it was found that the correct amplitude is not given by the
formula (\ref{phidot}), which is divergent, but by  (\ref{vprime}).
It has been further argued recently \cite{misao} that the formula (\ref{vprime})
may continue to be a useful approximation even when evolution of $\phi$
is not  described by a slow-roll solution. 
The other is the case $V'[\phi]$ vanishes, namely, when the inflaton
crosses over an extremum of the potential.  
In this case (\ref{vprime}) is divergent and would not be applicable.
The purpose of the present paper is first to analyze density fluctuation in
this situation and then extend it to the generation of perturbations in more
general non-slow-roll inflation models whose potential satisfies several
conditions indicated below to allow our analytic treatment.

First, as a specific model we consider the case $\phi$ overshoots the local
maximum of the potential at $\phi=0$ during the new
inflation stage in the chaotic new inflation model \cite{JY}.  More
specifically, we approximate the potential near the extremum as
\beq
 V[\phi] = V_0 - \frac{1}{2}m^2\phi^2,~~~~m^2>0,
\eeq
and consider the case $\phi$ crosses over $\phi=0$ from the positive
side to the negative one with a small kinetic energy so that the cosmic
expansion rate $H$ is practically constant in the relevant period.
We assume $H^2 \gg m^2$ and $V_0 \gg m^2\phi^2$.
Then the solution of (\ref{eqm}) reads,
\beqa
\phi(t) &= \dot{\phi}_{\ast}\biggl[\frac{1}{\lambda_+ - \lambda_-}e^{\lambda_+(t-t_{\ast})}
- \frac{1}{\lambda_+ - \lambda_-}e^{\lambda_-(t-t_{\ast})} \biggr],
\label{eq:phifull} \\
\dot{\phi}(t) &= \dot{\phi}_{\ast}\biggl[\frac{\lambda_+}{\lambda_+ - \lambda_-}e^{\lambda_+(t-t_{\ast})}
- \frac{\lambda_-}{\lambda_+ - \lambda_-}e^{\lambda_-(t-t_{\ast})} \biggr],
\label{eq:phidotfull} \\
&\lambda_\pm = -\frac{3}{2}H \pm \sqrt{\frac{9}{4}H^2 + m^2}\,,
\eeqa
where we have set $\phi(t_{\ast})=0$ and $\phidot (t_{\ast})=\phidot_{\ast}$.
Using $\lambda_+\cong m^2/(3H)$ and $\lambda_-\cong -3H\lmk
1+\frac{m^2}{9H^2}\rmk$, we find 
\beqa
\phi(t) &\cong&
\frac{\dot{\phi}_{\ast}}{3H}\biggl[e^{\frac{m^2}{3H^2}H(t-t_{\ast})}
 - e^{-3H\lmk 1+\frac{m^2}{9H^2}\rmk (t-t_{\ast})} \biggr]
= \frac{\dot{\phi}_{\ast}}{3H}\biggl[e^{\frac{m^2}{3H^2}H(t-t_{\ast})}
 - e^{-\frac{m^2}{3H^2}H(t-t_{\ast})}\biggl(\frac{a(t_{\ast})}{a(t)}\biggr)^3\biggr],
\label{eq:phitop} \\
\dot{\phi}(t) &\cong& \dot{\phi}_{\ast}\biggl[\frac{m^2}{9H^2}e^{\frac{m^2}{3H^2}H(t-t_{\ast})} 
+ \biggl(1 - \frac{m^2}{9H^2}\biggr)
e^{-3H\lmk 1+\frac{m^2}{9H^2}\rmk (t-t_{\ast})} \biggr]
\label{eq:phidottop1} \nonumber \\
&=& \dot{\phi}_{\ast}\biggl[\frac{m^2}{9H^2}e^{\frac{m^2}{3H^2}H(t-t_{\ast})} 
+ \biggl(1 -\frac{m^2}{9H^2}\biggr)
e^{-\frac{m^2}{3H^2}H(t-t_{\ast})}\biggl(\frac{a(t_{\ast})}{a(t)}\biggr)^3\biggr] 
\equiv \phidots (t) + \phidotr (t), \label{phidotsr}\\
{\rm with}~~~~\phidots (t) &\equiv&
\dot{\phi}_{\ast}\frac{m^2}{9H^2}e^{\frac{m^2}{3H^2}H(t-t_{\ast})},
~~~\phidotr (t) \equiv \dot{\phi}_{\ast}\biggl(1 - \frac{m^2}{9H^2}\biggr)
e^{-\frac{m^2}{3H^2}H(t-t_{\ast})}
\biggl(\frac{a(t_{\ast})}{a(t)}\biggr)^3,  \label{eq:phidottop}
\eeqa
near the local maximum of the potential.  Here $a(t)\propto e^{Ht}$
denotes the scale factor.  As is seen here $\dot{\phi}$ can be
decomposed into a slowly varying mode $\phidots (t)$ and a rapidly changing 
decaying mode $\phidotr (t)$ which is approximately
proportional to $a^{-3}(t) \propto e^{-3Ht}$.

We now incorporate  linear metric perturbations to the spatially flat
Robertson-Walker background 
in the longitudinal
gauge,
\begin{equation}
ds^2
= -\lkk 1+2\Psi(\textbf{x},t)\rkk dt^2
+a^2(t)\lkk 1+2\Phi(\textbf{x},t)\rkk d\textbf{x}^2,
\end{equation}
where we use the notation of~\cite{KS} for the gauge-invariant 
perturbation variables \cite{Baa}.
Hereafter all perturbation variables represent Fourier modes like
\begin{equation}
\Phi_\bfk=\int
\frac{d^3 x}{(2\pi)^{3/2}}\Phi(\textbf{x},t)e^{i\textbf{kx}},
\end{equation}
and we omit the suffix $\bfk$.
We use the following combination of  gauge-invariant variables 
\cite{sasaki,Mu,oseto,KH,NT},
\begin{equation}
Y=X-\frac{\dot{\phi}}{H}\Phi,
\label{defY}
\end{equation}
where
\begin{equation}
X=\delta\phi-\frac{a}{k}\dot{\phi}\sigma_g,
\end{equation}
is the gauge-invariant scalar field fluctuation
with $\sigma_g$ being the shear of each constant time
slice. The latter vanishes on Newtonian slice including the longitudinal
gauge.

The above quantity $Y$ is related to the gauge-invariant variable $\rc$ as
\begin{equation}
  \rc = \Phi-\frac{aH}{k}v=\Phi
  + \frac{2}{3}\frac{\Phi+H^{-1}\dot{\Phi}}{1+w}=-\frac{H}{\dot{\phi}}Y,
\end{equation}
in the present situation where the universe is dominated by the scalar field.
Here $v$ is a gauge-invariant velocity perturbation \cite{KS} and $w$
denotes the ratio of pressure to energy density.


From the perturbed Einstein equations, we obtain the following equations.
\begin{eqnarray}
\dot{X}\dot{\phi}-\ddot{\phi}X+\dot{\phi}^2\Phi=\frac{2}{\kappa^2}\frac{k^2}
{a^2}\Phi,
\\
\dot{\Phi}+H\Phi=-\frac{\kappa^2}{2}\dot{\phi}X,
\end{eqnarray}
where $\kappa^2=8\pi G$.
From these equations and (\ref{defY}), the equation of motion of $Y$ reads
\begin{equation}
\ddot{Y}+3H\dot{Y}+\left[\left(\frac{k}{a}\right)^2+ M_{Y \rm eff}^2
  \right]Y=0,
\label{eom.Y}
\end{equation}
with
\beq
  M_{Y \rm eff}^2 \equiv V''(\phi)+3\kappa^2\dot{\phi}^
2-\frac{\kappa^4}{2H^2}\dot{\phi}^4+2\kappa^2\frac{\dot{\phi}}{H}V'(\phi).
\label{MYeff:def}\eeq
This equation has the following exact solution in the long-wavelength
limit $k \longrightarrow 0$ ~\cite{KH,H}.
\begin{eqnarray}
Y(t) &=& {c}_1(k)Y_1(t)+{c}_2(k)k^3 Y_2(t), \label{exsol} \\
Y_1(t) &=& \frac{\dot{\phi}}{H}, \\
Y_2(t) &=& \frac{\dot{\phi}}{H}\int^t_{t_i}\frac{H^2}{a^3\dot{\phi}^2}dt,
\end{eqnarray}
where ${c}_1(k)$ and ${c}_2(k)$
are integration constants to be determined by
quantum fluctuations generated during inflation, and $t_i$ is some
initial time which may be chosen arbitrarily
because its effect can be absorbed by 
redefining  ${c}_1(k)$.  We choose $t_i$ so that $Y_2$ consists only of
the decaying mode.
Despite its appearance, the decaying mode $Y_2(t)$ is 
regular even at the epochs
$\dot{\phi}=0$ \cite{KH}.  In the second term of the right-hand-side of
(\ref{exsol}), an extra factor $k^3$ has been 
introduced so that the scale factor $a(t)$ appears in the rescaling-invariant form of
$k/a(t)$, and $c_1(k)$ and $c_2(k)$ have the same
dimension \cite{oseto}.

Using (\ref{eq:phidottop}) we find
\beqa
 Y_1(t) &=&  \frac{\phidots (t)}{H}+\frac{\phidotr (t)}{H}=
\frac{\dot{\phi}_{\ast}}{H}\biggl[\frac{m^2}{9H^2}e^{\frac{m^2}{3H^2}H(t-t_{\ast})} 
 + \biggl(1 - \frac{m^2}{9H^2}\biggr)
e^{-\frac{m^2}{3H^2}H(t-t_{\ast})}
\biggl(\frac{a(t_{\ast})}{a(t)}\biggr)^3\biggr], \\
Y_2(t) &=& -\frac{1}{3\phidots (t)a^3(t)}
\lmk 1+\frac{2m^2}{9H^2}\rmk^{-1}
=-\frac{3H^2}{m^2\dot{\phi}_{\ast}}
\lmk 1+\frac{2m^2}{9H^2}\rmk^{-1}
e^{-\frac{m^2}{3H^2}H(t-t_{\ast})}\frac{1}{a^3(t)},
\eeqa
so that the solution in the long wavelength limit is given by
\beqa
Y(t) &=& c_1(k)\lmk \frac{\phidots (t)}{H}+\frac{\phidotr (t)}{H}\rmk
- c_2(k)\frac{H^3}{3\phidots (t)}\lmk 1+\frac{2m^2}{9H^2}\rmk^{-1}
\biggl(\frac{k}{a(t)H}\biggr)^3 \label{yexact}
\eeqa
We can also incorporate the finite wavenumber correction using the
iterative expression \cite{oseto,KH} as
\begin{equation}
Y = c_1(k)Y_1+c_2(k)k^3Y_2+k^2Y_1\int aY_2Ydt-k^2Y_2\int aY_1Ydt. 
\end{equation}
In the present case we find the leading finite-wavenumber correction is
proportional to $\lmk k/a(t)\rmk^2$.  As a result we obtain
\beqa
Y(t) &=& c_1(k)\frac{\phidots (t)}{H}+ c_1(k)\frac{\phidots (t)}{2H}
\lmk 1-\frac{2m^2}{3H^2}\rmk 
\lmk\frac{k}{a(t)H}\rmk^2
\nonumber \\
&~&
+c_1(k)\frac{\phidotr (t)}{H}  
- c_2(k)\frac{H^3}{3m^2\phidots (t)}
\lmk 1-\frac{2m^2}{9H^2}\rmk
\biggl(\frac{k}{a(t)H}\biggr)^3+\cdots , \label{ykai}
\eeqa
to the lowest-order in $m^2/H^2$.
\if
This formula is valid in general as long as $Y_1$ consists of a slowly
changing non-decreasing mode and a decaying mode proportional to
$a^{-3}$ and $Y_2$ also decays in proportion to $a^{-3}(t)$.
In the model under consideration it is explicitly given by
\beqa
Y(t) &=& c_1(k)\biggl[\frac{m^2\dot{\phi}_{\ast}}{9H^3} 
+ \frac{m^2\dot{\phi}_{\ast}}{18H^3}\biggl(\frac{k}{a(t)H}\biggr)^2
+ \frac{\dot{\phi}_{\ast}}{H}e^{-\frac{m^2}{3H^2}H(t-t_{\ast})}
\biggl(\frac{a(t_{\ast})}{a(t)}\biggr)^3 \biggr] \nonumber \\
&&- c_2(k)\frac{3H^5}{m^2\dot{\phi}_{\ast}}\biggl(1 - \frac{m^2}{9H^2}\biggr)
e^{-\frac{m^2}{3H^2}H(t-t_{\ast})}
\biggl(\frac{k}{a(t)H}\biggr)^3
+ \order\biggl(\biggl(\frac{k}{a(t)H}\biggr)^5\biggr).  \label{ym} 
\eeqa
\fi

Next we consider  evolution of $Y$ in its short-wavelength regime 
in order to set the initial condition out of quantum fluctuations.
Since we are primarily interested in $k$-mode that leaves the horizon when the inflaton
crosses over its potential extremum at $t=t_{\ast}$,
we only need to consider the evolution of $Y$ 
 when $m^2|t-t_{\ast}|/3H \ll 1$ holds during inflation.
Thus in this regime (\ref{eom.Y}) becomes
\begin{equation}
\ddot{Y}+3H\dot{Y}+\left(\frac{k}{a}\right)^2Y = 0.
\end{equation}
Under the condition $|\dot H/H^2|\ll1$ the solution of this equation 
satisfying the normalization condition, 
\begin{equation}
Y\dot{Y}^* -\dot Y Y^* = \frac{i}{a^3},
\end{equation}
is approximately given by
\begin{equation}
Y=\frac{iH}{\sqrt{2k^3}}\left[\alpha_\bfk (1+ik\eta)e^{-ik\eta}
-\beta_\bfk (1-ik\eta)e^{ik\eta}\right],
\end{equation}
where $\eta=-1/(Ha)$ is a conformal time, and $\alpha_\bfk$ and $\beta_\bfk$ are 
constants which satisfy
$
|\alpha_\bfk|^2-|\beta_\bfk|^2 = 1.
$
We shall choose $(\alpha_\bfk \,,\beta_\bfk ) = (1,0)$ so that the 
vacuum reduces to the one in Minkowski spacetime at the 
short-wavelength limit ($-k\eta \rightarrow\infty$).  We therefore 
obtain
\begin{equation}
Y=\frac{iH}{\sqrt{2k^3}}(1+ik\eta)e^{-ik\eta}=\frac{iH}{\sqrt{2k^3}}
\lkk 1+ \frac{1}{2}\lmk\frac{k}{a(t)H}\rmk^2+\frac{i}{3}\lmk\frac{k}{a(t)H}\rmk^3
+...\rkk,
\label{ksol}
\end{equation}
where the latter expansion is useful after $k$-mode has gone out of the
horizon in the period $t<t_{\ast}+3H^2/m^2$.

From (\ref{ksol}) and (\ref{ykai}), we find that under the condition $m^2
\ll H^2$ we can match the first and the second
terms of (\ref{ykai}) with those of (\ref{ksol}) and the third and the
fourth terms of (\ref{ykai}) with the third term of (\ref{ksol}).
The former constitutes the growing (or more properly, nondecreasing)
mode and the latter corresponds to the decaying mode.  As a result we find
\beqa
c_1(k) &=& \frac{iH^2}{\sqrt{2k^3}\phidots (t_k)}=
\frac{9iH^4}{\sqrt{2k^3}m^2\dot{\phi}_{\ast}}e^{-\frac{m^2}{3H^2}H(t_k-t_{\ast})}, 
\label{33} \\
c_2(k) &=& 
\frac{\phidots (t_k)}{\sqrt{2k^3}H^2} \lmk 1+\frac{2m^2}{9H^2}\rmk
\biggl[ 1 + i\frac{27H^2}{m^2}\biggl(1 -
\frac{m^2}{9H^2}\biggr)
e^{-\frac{2m^2}{3H^2}H(t_k-t_{\ast})}
\biggl(\frac{a(t_{\ast})H}{k}\biggr)^3\biggr]. \label{34}
\eeqa
This means that the nondecaying mode of 
the curvature perturbation on the comoving horizon scale
at $t=t_{\ast}$, $r_{\ast}$, is given by,
\beq
\Rc(r_{\ast})= \lkk \left| c_1(k)\right|^2\frac{4\pi k^3}{(2\pi)^3}\rkk^{1/2}=
 \frac{9H^4}{2\pi m^2|\dot{\phi}_{\ast}|} .  \label{rctop}
\eeq

Apparently it is enhanced by a factor $9H^2/m^2$ compared with
(\ref{phidot}).  This does not mean, however, that we find a severe
deviation from a scale-invariant spectrum.
On a more general comoving scale $r=r(t_k)=2\pi/k$ leaving the Hubble radius during
new inflation, we find from the first equatlity of (\ref{33}),
\beq
\Rc(r)= \frac{H^2}{2\pi |\phidots (t_k)|},
\label{rcgeneral}
\eeq
which is practically scale-invariant for $m^2 \ll H^2$.  The deviation
from the scale-invariant spectrum is manifest only in the decaying mode
$c_2(k)$, as is seen in the second term of (\ref{34}). 
\if
In the above expression, the second factor, which is the correction
factor to the slow-roll formula (\ref{phidot}), 
is equal to $9H^2/m^2$ for the mode leaving the Hubble radius at
$t_k=t_{\ast}$ but it rapidly decreases to unity as we focus on the shorter
length scales whose horizon-crossing epoch  
$t_k$ is larger than $t_{\ast}$.  Thus (\ref{rcgeneral}) 
approaches the formula (\ref{phidot}) for these modes.  Meanwhile $\phidot (t)$ also
changes with the same rate 
because the time dependence of the denominator in (\ref{rcgeneral})
is the same as that of the numerator as seen in (\ref{eq:phidottop}).  
We therefore find (\ref{rcgeneral}) is a
constant independent of the scale.  
\fi

Thus we can obtain the correct expression for the
nondecreasing mode of the curvature
perturbation if we replace $\phidot$ in (\ref{phidot}) with its
nondecaying mode only.  This was also the case when $\phidot$ vanishes
during inflation which was studied in \cite{oseto}.
Figure 1 depicts the actual spectrum of curvature
perturbation together with the
conventional slow-roll formula (\ref{phidot}) for $H^2/m^2=30$. 
Using (\ref{phidot}) would seriously underestimate the actual amplitude
in the regime $\phi$ is not slowly rolling.

\begin{figure}
  \begin{center}
    \centerline{\psfig{figure=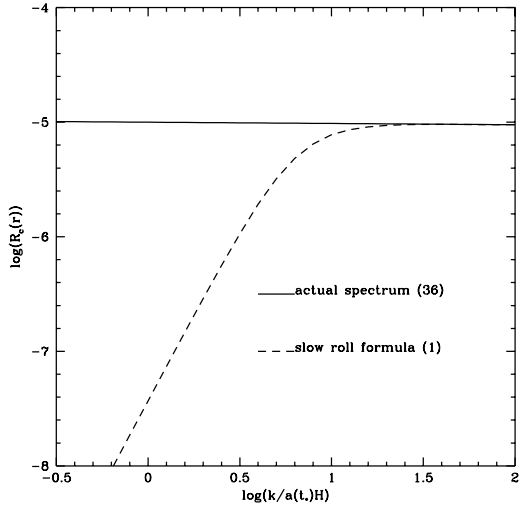,width=8cm}}
      \caption{
      Actual spectrum of the curvature perturbation (solid line) and the result of
   slow-roll formula (broken line) for $H^2/m^2=30$.}
      \label{fig:nch}
  \end{center}
\end{figure}

\if
The fact that (\ref{rcgeneral}) is constant and that it approaches the
slow-roll formula in the limit $\phidot$ contains only the nondecaying
mode means that we can obtain the correct expression for the curvature
perturbation if we replace $\phidot$ in (\ref{phidot}) with its
nondecaying mode only.  This was also the case when $\phidot$ vanishes
during inflation which was studied in \cite{oseto}.

In conclusion, neither (\ref{phidot}) nor (\ref{vprime}) gives correct
amplitude of curvature perturbation in general non-slowly rolling
inflation models.  We should use (\ref{phidot}) with $\phidot$ replaced
by its nondecaying component only.
\fi

Next we consider the applicability of this approach to a more general
potential $V[\phi]$.  As is clear from the above analysis our method is
applicable if the solution of $\phidot$ is separable to a sum of a 
slowly varying mode and a decaying mode approximately proportional to
$a^{-3}(t)$ at least for a few expansion time scales after the $k$-mode
under consideration leaves the Hubble radius at $t=t_k$. 
Let us expand $V[\phi]$ around $\phi=\phi(t_k)\equiv \phi_k$ up to
second order:
\beq
 V[\phi]=V[\phik]+V'[\phik](\phi-\phik)+\frac{1}{2}V''[\phik](\phi-\phik)^2+\cdots
 \equiv V_k+V'_k(\phi-\phik)+\frac{1}{2}V''_k(\phi-\phik)^2+\cdots .  \label{expand}
\eeq
Then a solution of the equation of motion, $\ddot{\phi}+3H_k\phidot
+\vonek +\vtwok(\phi-\phik)=0$, is given by a linear combination of 
$e^{\Lambda_+(t-t_k)}$ and $e^{\Lambda_-(t-t_k)}$, where $\Lambda_{\pm}$
are solutions of $\Lambda^2+3\hk \Lambda +\vtwok=0$.  Here a subscript
$k$ denotes the value of each quantity at the time $t_k$.

In order that the solution of $\phidot$ is separable to a slowly
changing mode and a decaying mode, we require $|\vtwok|\ll \hk^2$.  Then
we find, to the lowest order in $|\vtwok|/\hk^2$,
\beq
 \phidot(t)=\lkk \frac{\vtwok}{9\hk^2}\phidotk
+\lmk 1-\frac{\vtwok}{9\hk^2}\rmk\frac{\vonek}{3\hk}\rkk 
e^{\frac{\vtwok}{3\hk^2}\hk(t-t_k)}
+\lmk 1-\frac{\vtwok}{9\hk^2}\rmk\lmk \phidotk -\frac{\vonek}{3\hk}\rmk
e^{-3\lmk 1+\frac{\vtwok}{9\hk^2}\rmk\hk(t-t_k)}.  \label{phidotgeneral}
\eeq
The first and the second terms of the above solution correspond to 
$\phidots (t)$ and $\phidotr (t)$ in (\ref{phidotsr}) of the previous
analysis, respectively.  If we apply the same matching method as before, we obtain
\beq
\Rc(r)= \frac{\hk^2}{2\pi}\left| \frac{\vtwok}{9\hk^2}\phidotk
+\lmk 1-\frac{\vtwok}{9\hk^2}\rmk\frac{\vonek}{3\hk}\right|^{-1}, \label{rcmoregeneral}
\eeq 
instead of (\ref{rcgeneral}) for the nondecaying mode of the curvature
perturbation on the comoving horizon scale at $t=t_k$.  In order that 
(\ref{rcmoregeneral}) gives the correct expression for the curvature
perturbation, the expansion (\ref{expand}) should remain valid at least
for a few expansion time scales after $k$-mode has crossed the horizon
at $t=t_k$.  That is, we require
\beq
  |\vonek||\phidotk|\hk^{-1} < V_k,~~~~~|\vtwok|(\phidotk\hk^{-1})^2
< V_k.
\eeq
Since the $k$-mode leaves the horizon during inflation, we have
$\phidotk^2 < V_k$.  Then together with the assumption $|\vtwok| \ll
\hk^2$, the second inequality is trivially satisfied.  Furthermore,
since $\vonek/\hk$ is of the same order of $\phidot$ in the slow-roll
case, we find $|\vonek||\phidotk|\hk^{-1} \lesssim  \phidotk^2
< V_k$ is also automatically satisfied.  Thus the only
nontrivial condition for our analysis to be valid is $|\vtwok| \ll
\hk^2$ apart from the condition for accelerated expansion $\phidotk^2 < V_k$.

In summary, we have analytically studied generation of curvature perturbation in the
case the slow-roll equation of motion does not hold during inflation, by
matching the quantum fluctuation (\ref{ksol}) with the long-wave exact solution
(\ref{yexact}).  For this matching to be possible it is essential that the
solution of $\phidot$ can be described by a sum of a slowly changing
mode and a decaying mode approximately
 proportional to $a^{-3}(t)$, which requires $|\vtwok|\ll
\hk^2$.  Models with more complicated potentials that violate this
condition must be studied numerically \cite{misao,num}. 

\acknowledgements
{
We are grateful to M.\ Sasaki and O.\ Seto for useful communications.
The work of JY was supported in part by the Monbukagakusho Grant-in-Aid for Scientific
Research
 ``Priority Area: Supersymmetry and
Unified Theory
of Elementary Particles (No.\ 707).''}

\end{document}